\documentclass[twocolumn, superscriptaddress,preprintnumbers,amsmath,amssymb]{revtex4-1}

\usepackage{graphicx}
\usepackage{dcolumn}
\usepackage{bm}
\usepackage{mathtools}
\usepackage{epstopdf}
\usepackage{float}

\begin{document}

\preprint{Version: \today}

\title{Coherent Oscillations of Driven rf SQUID Metamaterials}

\author{Melissa Trepanier}
\affiliation{Department of Physics, CNAM, University of Maryland, College Park, MD 20742, USA}
\author{Daimeng Zhang}
\affiliation{Department of Electrical and Computer Engineering, University of Maryland, College Park, MD 20742, USA}
\author{Oleg Mukhanov}
\affiliation{Hypres, Inc., 175 Clearbrook Road, Elmsford, NY 10523, USA}
\author{V. P. Koshelets}
\affiliation{Laboratory of Superconducting Devices for Signal Detection and Processing, KotelÕnikov Institute of Radio Engineering and Electronics, Moscow 125009, Russia}
\author{Philipp Jung}
\author{Susanne Butz}
\affiliation{Physikalisches Institut, Karlsruhe Institute of Technology, Germany}
\author{Edward Ott}
\author{Thomas M. Antonsen}
\affiliation{Department of Physics, CNAM, University of Maryland, College Park, MD 20742, USA}
\affiliation{Department of Electrical and Computer Engineering, University of Maryland, College Park, MD 20742, USA}
\author{Alexey V. Ustinov}
\affiliation{Physikalisches Institut, Karlsruhe Institute of Technology, Germany}
\affiliation{Russian Quantum Center, National University of Science and Technology MISIS, Moscow 119049, Russia}
\author{Steven M. Anlage}
\affiliation{Department of Physics, CNAM, University of Maryland, College Park, MD 20742, USA}
\affiliation{Department of Electrical and Computer Engineering, University of Maryland, College Park, MD 20742, USA}
\date{\today}
\begin{abstract}
Through experiments and numerical simulations we explore the behavior of rf SQUID (radio frequency superconducting quantum interference device) metamaterials, which show extreme tunability and nonlinearity. The emergent electromagnetic properties of this metamaterial are sensitive to the degree of coherent response of the driven interacting SQUIDs. Coherence suffers in the presence of disorder, which is experimentally found to be mainly due to a dc flux gradient. We demonstrate methods to recover the coherence, specifically by varying the coupling between the SQUID meta-atoms and increasing the temperature or the amplitude of the applied rf flux.
\end{abstract}
\maketitle

Metamaterials are artificially structured media with electromagnetic properties arising from the structure of individual meta-atoms and the interactions between them. Metamaterials can have emergent properties not seen in natural materials e.g. a negative index of refraction \cite{Veselago2003, Smith2000, Shelby2001}, cloaking \cite{Alu2003, Schurig2006}, and super-resolution imaging \cite{Pendry2000, Jacob2006}. 
Collections of superconducting split ring resonators (SRRs) have an effective permeability that can be tuned by suppressing superconductivity with increased temperature and applied magnetic field \cite{Ricci2005, Ricci2006, Ricci2007a, Fedotov2010OE}, or applied current \cite{Savinov2012}. Suppressing superconductivity tunes the kinetic inductance but this process increases losses and can be slow.

The rf SQUID, which has a Josephson junction instead of the capacitive gap, is a significant improvement over the SRR; by applying a magnetic field the self-resonance can be tuned quickly over a wide range without a substantial increase in losses \cite{Trepanier2013}. Using an rf SQUID as a meta-atom was proposed theoretically \cite{Du2006, Lazarides2007, Caputo2012} and experimentally demonstrated \cite{Jung2013, Trepanier2013}.
Previous experimental work on rf SQUID array metamaterials has been limited to 1D arrays  \cite{Butz2013, Butz2013a, Jung2014} and theoretical work has only considered nearest neighbor coupling between the SQUIDs. \cite{Lazarides2013, Lazarides2010, Tsironis2014, Lazarides2015a, Lazarides2015}. In this paper, we consider dense globally coupled 2D arrays and study the behavior resulting from the complex interactions between the SQUIDs, not seen in a 1D configuration.

One of the challenges of nonlinear metamaterials is understanding and controlling their collective behavior, which is not a simple linear superposition of the response of each meta-atom. An rf SQUID metamaterial is an array of driven linearly-coupled nonlinear oscillators \cite{Zhang2016}. The Kuramoto model has been used to study coherence in related systems,  such as 1D arrays of current-biased Josephson junctions \cite{Tsang1991, Acebron2005, Marvel2009}. The typical Kuramoto system is a collection of linear harmonic oscillators with a Gaussian distribution of self-resonant frequencies. These oscillators interact through nonlinear uniform all-to-all coupling. Under certain conditions the entire array can oscillate in phase at the same frequency (coherence), despite the differences in self-resonant frequencies.

The Kuramoto model quantifies coherence with an order parameter, $r=\left| \frac{1}{N}\sum_j^N e^{i\theta_j}\right|$
where $\theta_j$ is the phase of the $j$th oscillator and $N$ is the number of oscillators. Perfect coherence ($r=1$) is achieved when the SQUIDs are all oscillating in phase at the same frequency. The Kuramoto model order parameter has been used to quantify coherence in numerical studies of 1D rf SQUID arrays \cite{Lazarides2013}, but we find the modified order parameter presented in this paper more useful. The concept of coherence has also been explored in the context of other metamaterials, specifically an ASR (asymmetric split ring) array \cite{Fedotov2010, Jenkins2012, Jenkins2013}.

Coherence has consequences for the performance of the SQUID array as a metamaterial. Coherence is suppressed in the experiments discussed below when different SQUIDs in the array experience different applied dc magnetic flux. This occurs despite extensive magnetic shielding because of the SQUID's extreme sensitivity, (properties can change significantly even for field variations smaller than 1 $\mu$T). This paper presents tactics to minimize the effects of this disorder on the emergent properties of the metamaterial.

The remainder of the paper is organized as follows. First there is an explanation of the methods for numerically simulating and experimentally measuring the resonant response and coherence of the SQUID metamaterials. Then we present experimental and simulation results that explore how the coherence is suppressed in the experiment by non-uniform dc flux bias. This is followed by a discussion of how the coherence affects the performance of the rf SQUID array as a metamaterial and how the coherence can be recovered with increased coupling, rf driving flux, and temperature. The paper closes with a summary and conclusions.

\begin{figure}[t]
\includegraphics*[width=75mm]{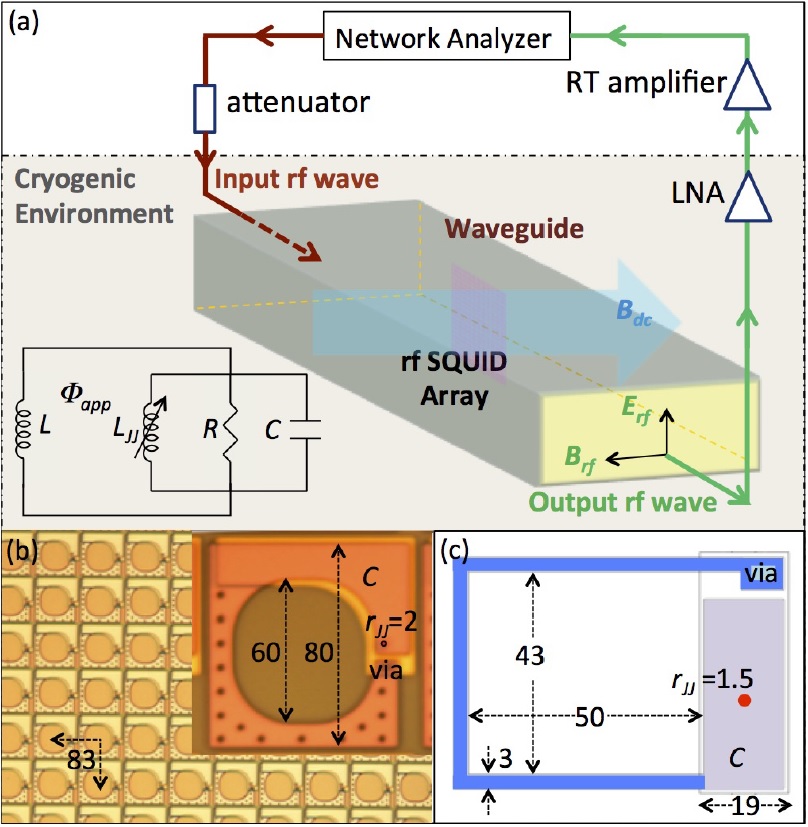}
\caption{(a) Experimental setup of waveguide transmission measurement with inset showing schematic of an rf SQUID including the RCSJ model for the Josephson junction. (b) Portion of 27x27 array along with SQUID design with lengths given in $\mu$m. (c) Design of the IREE SQUID from the 21x21 array with lengths given in $\mu$m and $r_{JJ}$ denoting the radius of the Josephson junction.}
\label{fig1}
\end{figure}

\textit{Modeling and simulations} Using the resistively and capacitively shunted junction (RCSJ) model the SQUID can be modeled by the equivalent circuit shown in Fig. \ref{fig1}(a). An array of $N$ coupled rf SQUIDs can be described by the following set of coupled nonlinear differential equations:
\begin{equation}
\frac{2\pi}{\Phi_0} (\hat{\Phi}_{dc}+\hat{\Phi}_{rf}\sin{\Omega \tau})=\hat{\delta}+\bar{\bar{\kappa}}(\beta_{rf}\sin{\hat{\delta}}+\gamma\hat{\delta}'+\hat{\delta}'')
\label{fullnl}
\end{equation}
where $\hat{\delta}$ is a vector of length $N$ representing the gauge-invariant phase difference across the junction for each of the $N$ SQUIDs. $\beta_{rf}=\frac{2\pi L I_c}{\Phi_0}$, $\gamma=\frac{1}{R} \sqrt{\frac{L}{C}}$, $\Omega=\frac{\omega}{\omega_{geo}}$, $\omega$ is the driving rf flux frequency, $\tau=t\omega_{geo}$, $\omega_{geo}=\frac{1}{\sqrt{LC}}$, and the prime denotes a derivative with respect to $\tau$.
$\Phi_0=h/2e$ is the flux quantum, $I_c$ is the critical current of the junction, $L$ is the self-inductance of the rf SQUID loop, $R$ is the resistance and $C$ is the capacitance of the junction, and $t$ is time.
$\hat{\Phi}_{dc}$ and $\hat{\Phi}_{rf}$ are vectors representing the the dc and rf applied flux in each SQUID, respectively. $\bar{\bar{\kappa}}$ is a $N$x$N$ 2D coupling matrix $\kappa_{ij}= \left\{
  \begin{array}{lr}
    1 &  : i=j\\
    M_{ij}/L &  : i \neq j
  \end{array}
\right.$ where $M_{ij}$ is the mutual inductance between SQUIDs $i$ and $j$. The off-diagonal elements are negative, because the coupling field created by one SQUID induces a diamagnetic response in its neighbor for the coplanar geometry used here. The coupling exists between every pair of SQUIDs.  Equation \ref{fullnl} can be solved for $\hat{\delta}(t)$ which can then be used to calculate any quantity of interest, such as microwave transmission through the metamaterial $S_{21}$ and $r_A$, a modified Kuramoto order parameter presented below. (For details on how Eq. \ref{fullnl} is solved including how the parameter values are chosen see the supplemental material.)

Although there are several key differences between a 2D array of rf SQUIDs and the basic Kuramoto system, a modified Kuramoto order parameter is still useful for quantifying coherence. The most important difference (i.e. the one that motivates the modification to the order parameter) is that the Kuramoto model assumes that the amplitudes of all the oscillators are the same and so the order parameter only considers phase information. Simulations of rf SQUID arrays show that the amplitudes of the gauge-invariant phase oscillations can have a wide distribution, for example between the middle and edge of the array. Consequently, a modified coherence order parameter that gives greater weight to the phase of oscillators with greater amplitude, and whose magnitude is still normalized to fall between 0 and 1, can be introduced as,
\begin{equation}
r_A e^{i\phi_A}= \frac{\sum_j^N A_j e^{i\theta_j}}{\sum_j^N A_j},
\label{r}
\end{equation}
where $A_j$ is the (real) amplitude of oscillation of $\delta_j(t)$ for the $j$th rf SQUID; the solutions for $\hat{\delta}(t)$ are harmonic to a very good approximation.

\textit{Experimental Setup} The 2D SQUID array (Fig. \ref{fig1}(b)) is oriented in a copper Ku rectangular waveguide so that the plane of the SQUIDs is perpendicular to the rf magnetic field of the traveling wave (Fig. \ref{fig1}(a)) similar to previous experiments \cite{Trepanier2013, Zhang2015, Zhang2016}. Superconducting coils surround the waveguide to apply an additional dc bias field. The transmission vs. rf flux driving frequency $S_{21}(\omega)$ is measured through the waveguide by a microwave network analyzer; the resonant response of the metamaterial appears as a dip in this curve.

This paper considers results from two rf SQUID arrays. The SQUIDs are composed Nb loops with AlO$_x$ junctions on silicon substrates and are all non-hysteretic ($\beta_{rf}<1$). One of the arrays is a 21x21 array prepared by IREE \cite{Koshelets1989, Koshelets1991, Filippenko2001}. The coupling between nearest neighbors $\kappa_0=-0.02$ and $L=0.13$ nH are calculated numerically with Fasthenry \cite{Whiteley}. Fasthenry calculations show that the maximum coupling of SQUIDs with this design in a rectangular array is $\kappa_0=-0.06$ (when the SQUIDs are as close together as possible). The parameter values are as follows (Fig. \ref{fig1}(c)): $C=2.1$ pF, $I_c=1.95$ $\mu$A ($\beta_{rf}=0.77$), and $R=1000$ $\Omega$. The other array is a 27x27 array prepared by Hypres and shown in Fig. \ref{fig1} (b) \cite{Yohannes2005, Yohannes2007, Tolpygo2010}. The parameter values are as follows: $\kappa_0=-0.03$, $L=0.13$ nH, $C=2.2$ pF, $I_c=2.2$ $\mu$A ($\beta_{rf}=0.87$), and $R=1500$ $\Omega$.

\begin{figure}[t]
\includegraphics*[width=75mm]{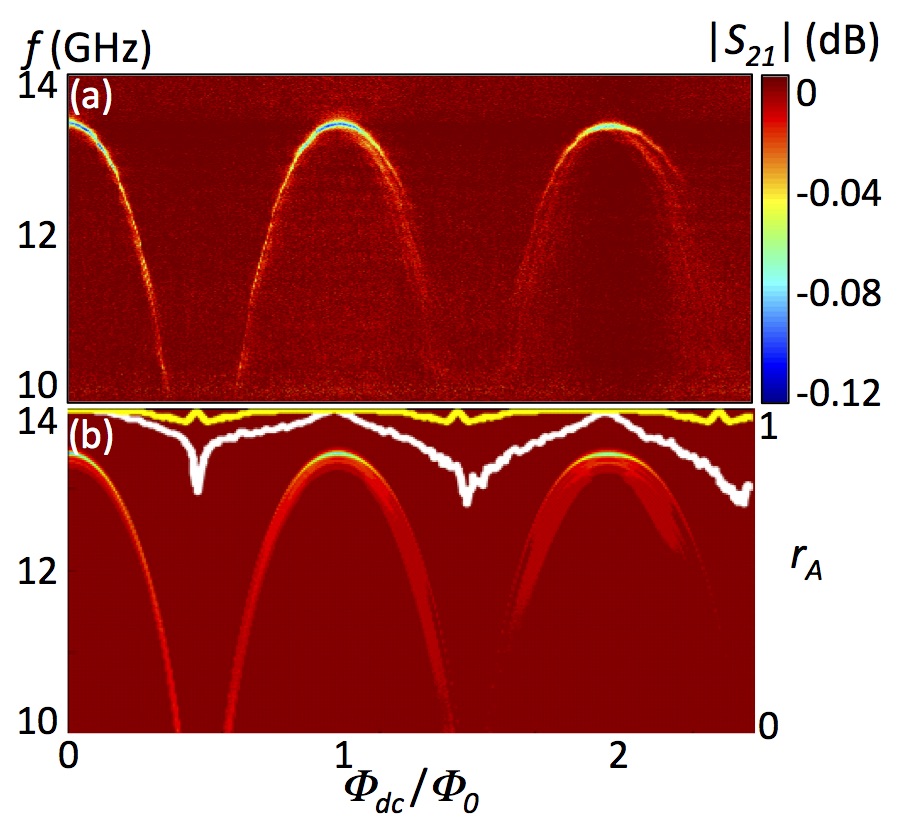}
\caption{(a) Measured (b) and simulated transmission for the 21x21 array with $|\kappa_0|=0.02$ as a function of frequency and dc flux in the limit $\Phi_{rf}/\Phi_0\ll 1$. The simulation has a dc flux gradient such that one edge of the array experiences the $\Phi_{dc}/\Phi_0$ value shown and the other edge is 90$\%$ of that value. Inset curves show simulated coherence $r_A$ as a function of applied dc flux with (white) and without (yellow) the flux gradient.}
\label{cohercoup}
\end{figure}

\textit{Results}
Fig. \ref{cohercoup}(a) shows the measured transmission vs. frequency and dc magnetic flux at low rf flux.  The metamaterial resonant response tunes considerably with dc flux.  In the absence of any disorder the resonant response of the array is periodic in dc flux. However, the experimental results shown in Fig. \ref{cohercoup} (a) do not have this periodicity. As dc flux increases the resonance dip becomes wider and shallower, the maximum resonant frequency (when $\Phi_{dc}/\Phi_0$ is an integer value) decreases by 0.04 GHz, and there is distinct splitting of the resonance dip.

These features are reproduced in the simulation Fig. \ref{cohercoup} (b) and explained by the model when a linear dc flux gradient is applied such that flux at one edge of the array is $90\%$ of that at the other. 
We have also considered in simulation other likely types of disorder: Gaussian-random distributions of coupling strength, dc flux, critical currents, and dissipation. None of these cause a progressive loss of coherence with increased dc flux seen in the experimental results.
For further experimental and simulation evidence that there is a dc flux gradient in the experiment causing a loss of coherence (and not other types of disorder) see the supplemental material.

\begin{figure}[h]
\includegraphics*[width=75mm]{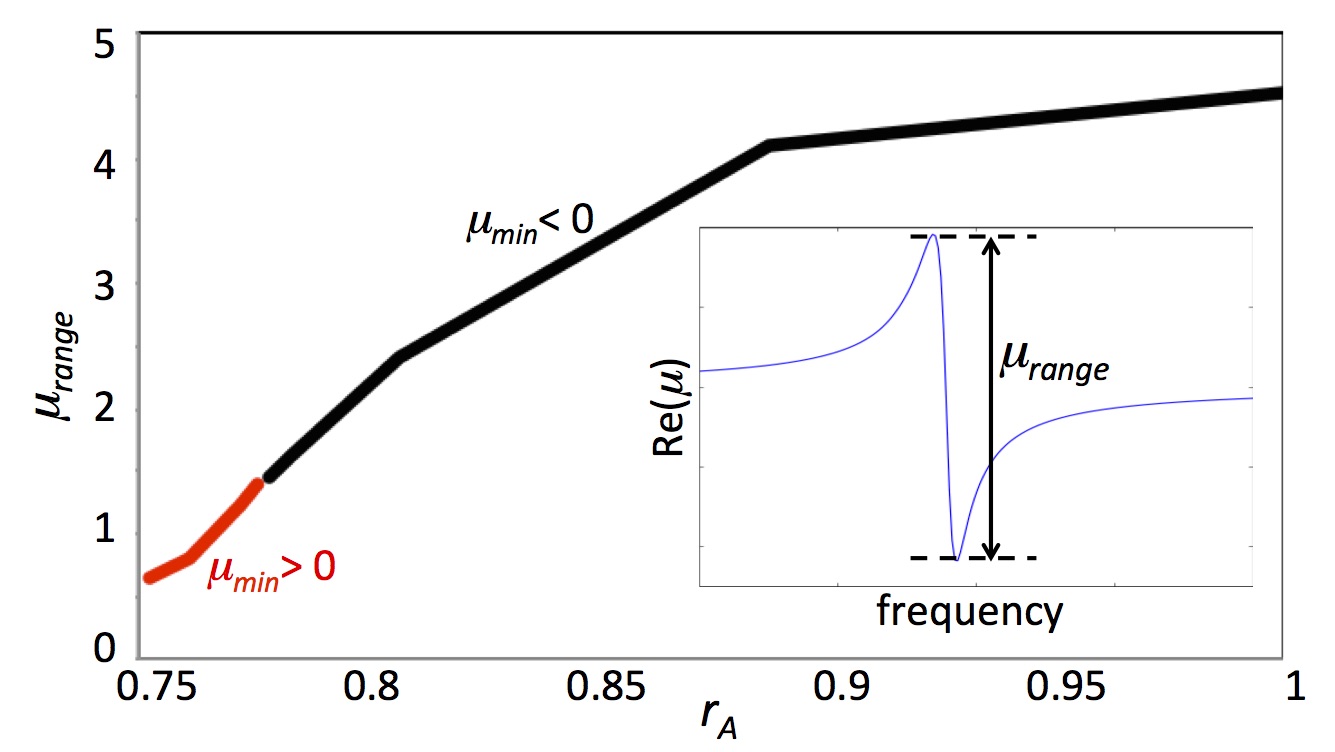}
\caption{Numerical simulation for the range of frequency tunability in the real part of the effective permeability as a function of coherence for eight non-interacting 21x21 arrays with $|\kappa_0|=0.02$. The coherence was varied by applying a dc flux gradient. The black portion of the curve is where the minimum effective permeability is negative. Inset: simulated real part of effective permeability as a function of frequency illustrating how the range of effective permeability is defined.}
\label{mu}
\end{figure}
To optimize the performance of the rf SQUIDs as a metamaterial, it is necessary to maximize the coherence $r_A$. A decrease in coherence (caused by an increase in the dc flux gradient) results in, among other things, a reduced range of tunability for the effective permeability. Figure \ref{mu} illustrates how this emergent electromagnetic property of the metamaterial can be improved by increasing the coherence, $r_A$.

\begin{figure}[t]
\includegraphics*[width=75mm]{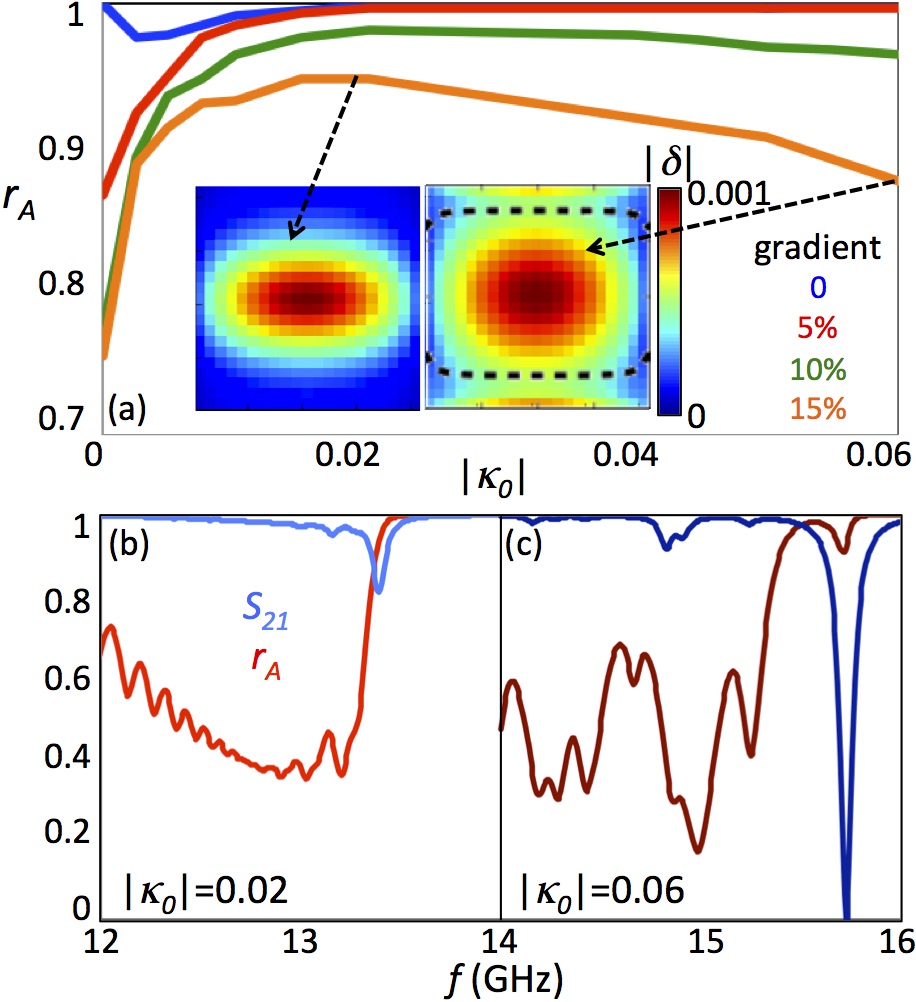}
\caption{(a) Numerically simulated coherence of a 21x21 rf SQUID array as a function of coupling on the primary resonance at $\Phi_{dc}/\Phi_0=2$ for three different flux gradients in the limit $\Phi_{rf}/\Phi_0\ll 1$. Insets: Simulated spatial distribution of amplitude (color) and phase (dashed contour line at $\theta_j=0$) of $\hat{\delta}(t)$ with a gradient such that one edge of the array has $15\%$ of the dc flux of the other edge. (b, c) Simulated transmission and coherence vs. frequency for two different coupling values.}
\label{cohercoup2}
\end{figure}

\textit{Re-establishing Coherence} According to numerical studies, one way to mitigate the effects of the flux gradient is to increase the coupling between the SQUIDs. When there is no coupling and the dc flux is uniform the oscillators are perfectly coherent $r_A=1$ with exactly the same amplitude and phase.
Increasing coupling ($|\kappa_{0}|$) causes an initial slight decrease in coherence as shown in the blue curve of Fig. \ref{cohercoup2} (a). With coupling the SQUIDs at the edge experience different flux from those at the center because they have fewer neighbors. Further increasing the coupling decreases the phase difference between the edges and the center which increases coherence until it saturates at $r_A=1$. 

The tendency for coupling to enhance the coherence persists in the presence of a dc flux gradient. Small amounts of coupling improve coherence regardless of the magnitude of the applied dc flux gradient, (see the low $|\kappa_0|$ part of Fig. \ref{cohercoup2} (a)). However, the coherence as a function of coupling saturates for small flux gradients and actually decreases for larger gradients.
This drop occurs because the increased coupling recruits additional SQUIDs to participate in the oscillation, but these SQUIDs are out of phase (an example is shown in the inset of Fig. \ref{cohercoup2} (a)) causing the metamaterial coherence to decrease. This suggests that there is an optimal value for the coupling in the presence of a dc flux gradient; for the 21x21 SQUID array this is about $|\kappa_0|=0.02$. 

\begin{figure}[t]
\includegraphics*[width=75mm]{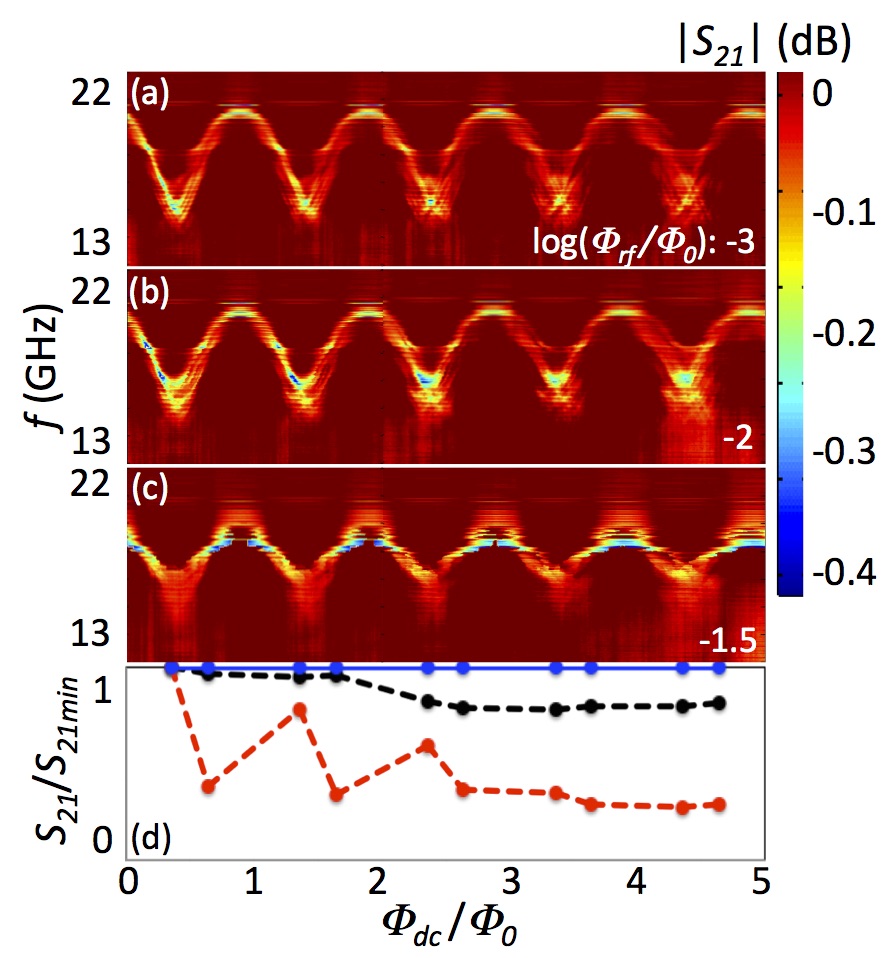}
\caption{(a-c) Measured transmission as a function of frequency and dc flux at three values of rf flux for 27x27 array $T=7$ K. (d) Local minima (normalized by the global minimum) in measured transmission as a function of dc flux $S_{21}(\Phi_{dc})$ at the geometric resonant frequency $\omega_{geo}/2\pi$ for (red) $\log_{10} (\Phi_{rf}/\Phi_0)=-3$, (black) $\log_{10} (\Phi_{rf}/\Phi_0)=-1.5$, (blue) if no flux gradient were present, independent of the rf flux.}
\label{trade}
\end{figure}
At higher flux gradients even when the coherence decreases with increasing coupling, the $S_{21}(\omega)$ dip continues to deepen, see Fig. \ref{cohercoup2} (b) and (c). This is because the depth of the dip in $S_{21}(\omega)$ depends on the sum of the amplitudes of $\hat{\delta}$ independent of the phase. 

Another method for mitigating the loss of coherence due to the dc flux gradient is to decrease the range of dc flux tunability, for example by increasing temperature or rf flux. The decrease in dc flux sensitivity makes the array less sensitive to dc flux disorder, improving coherence. Figure \ref{trade} shows experimentally how the coherence of the array is improved at higher rf flux; the symptoms of the coherence loss with increasing dc flux (\textit{i.e.} the $S_{21}(\omega)$ dip becomes broader, shallower, and splits and the maximum frequency decreases while the minimum frequency increases) are not as pronounced for higher rf flux values.

Of these symptoms the depth of the transmission dip is the easiest to quantify and is shown in Fig. \ref{trade} (d). If the array were coherent there would be no change in the depth of the $S_{21}(\Phi_{dc})$ dip with increased dc flux (blue curve). However, for low rf flux the dips become substantially shallower with increased dc flux indicating a loss of coherence (red curve). At the higher rf flux the transmission dips and coherence are significantly less affected by the dc flux gradient (black curve).

\begin{figure}[t]
\includegraphics*[width=75mm]{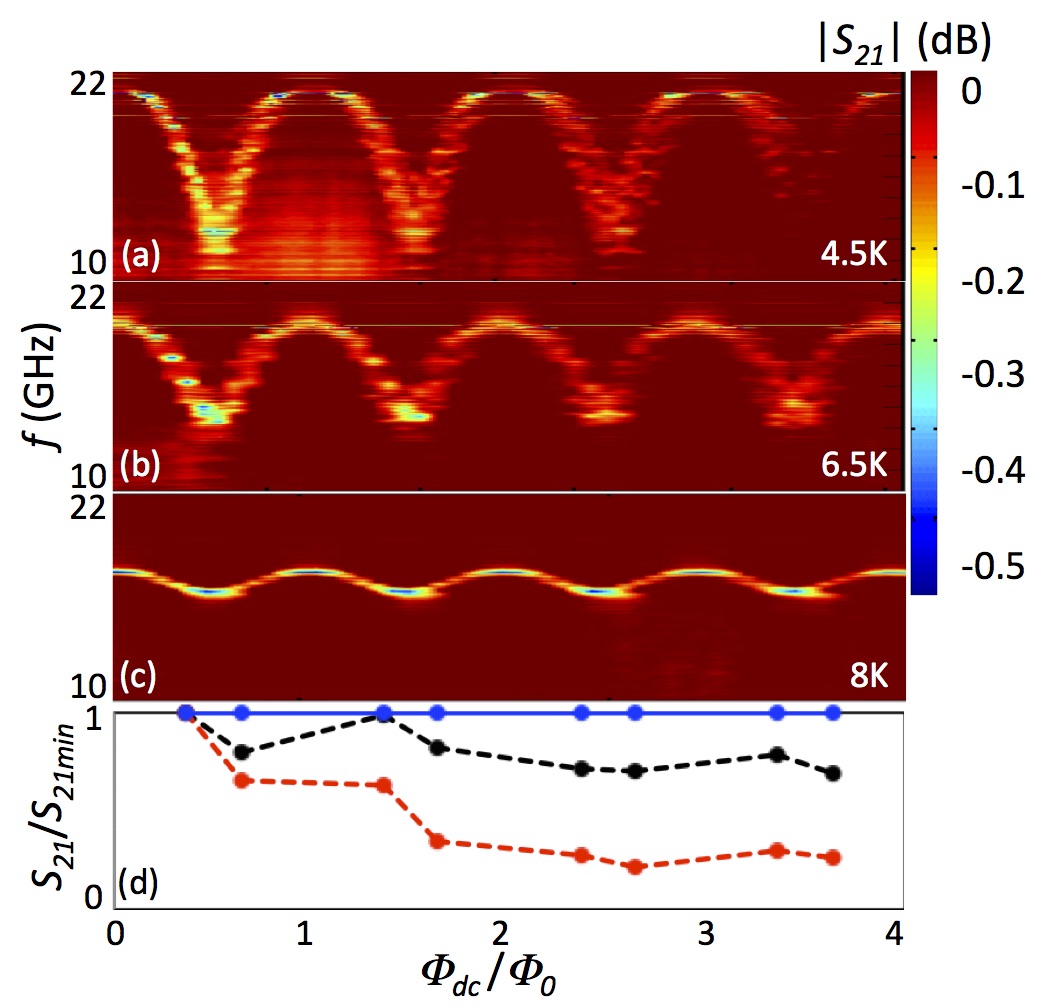}
\caption{(a-c) Measured transmission as a function of frequency and dc flux at three temperatures for the 27x27 array with $\log{\Phi_{rf}/\Phi_0}=-2$. (d) Local minima (normalized by the global minimum) in measured transmission as a function of dc flux $S_{21}(\Phi_{dc})$ at the geometric resonant frequency $\omega_{geo}/2\pi$ for (red) $T = 4.5 K$, (black) $T = 8.0 K$, (blue) if no flux gradient were present.}
\label{trade2}
\end{figure}

Increased temperature also decreases the dc flux sensitivity and improves coherence. Figure \ref{trade2} shows how the coherence is improved for higher temperatures (which brings about increased damping and reduced $\beta_{rf}$) just as it is for higher rf flux. The tradeoff between dc flux tuning and coherence can be adjusted after fabrication of the metamaterial through variation of the temperature and rf flux, unlike the coupling between SQUIDs which is determined by the array geometry.

\textit{Conclusions}
For a 2D array of rf SQUIDs to function as an effective medium with tunable permeability the SQUID meta-atoms must respond coherently to incident electromagnetic waves. In our experiment a dc flux gradient causes a substantial loss of coherence at higher dc flux values. This loss of coherence is evident in our measurements as a loss of periodicity in dc flux, a reduction in maximum resonant frequency, a broader and shallower resonant dip, and splitting of the resonance dip at higher dc flux values.

The coherence can be recovered by increasing coupling between the SQUIDs (up to a point, as shown numerically), or by decreasing their dc flux sensitivity \textit{i.e.} increasing rf flux or temperature. By using these strategies to maximize coherence and taking steps to minimize uneven dc flux bias, rf SQUID metamaterials can be tuned coherently over a broad frequency range. The large-magnitude, high-speed, low-loss tuning behavior that is observed in the single SQUID is also possible in a 2D SQUID metamaterial.

\textit{Acknowledgments}
This work is supported by the NSF-GOALI and OISE programs through grant $\#$ECCS-1158644, and the Maryland Center for Nanophysics and Advanced Materials (CNAM). This work was also supported in part by the Ministry of Education and Science of the Russian Federation in the framework of Increase Competitiveness Program of the NUST MISIS (contracts no. K2-2014-025, K2-2015-002, and K2-2016-063). The authors acknowledge the University of Maryland supercomputing resources (http://www.it.umd.edu/hpcc). We thank Masoud Radparvar and Georgy Prokopenko for helpful suggestions, Nikos Lazarides and George Tsironis for productive discussions, and H. J. Paik and M. V. Moody for use of the pulsed tube refrigerator.

\bibliography{Bib}

\begin{thebibliography}{40}%
\makeatletter
\providecommand \@ifxundefined [1]{%
 \@ifx{#1\undefined}
}%
\providecommand \@ifnum [1]{%
 \ifnum #1\expandafter \@firstoftwo
 \else \expandafter \@secondoftwo
 \fi
}%
\providecommand \@ifx [1]{%
 \ifx #1\expandafter \@firstoftwo
 \else \expandafter \@secondoftwo
 \fi
}%
\providecommand \natexlab [1]{#1}%
\providecommand \enquote  [1]{``#1''}%
\providecommand \bibnamefont  [1]{#1}%
\providecommand \bibfnamefont [1]{#1}%
\providecommand \citenamefont [1]{#1}%
\providecommand \href@noop [0]{\@secondoftwo}%
\providecommand \href [0]{\begingroup \@sanitize@url \@href}%
\providecommand \@href[1]{\@@startlink{#1}\@@href}%
\providecommand \@@href[1]{\endgroup#1\@@endlink}%
\providecommand \@sanitize@url [0]{\catcode `\\12\catcode `\$12\catcode
  `\&12\catcode `\#12\catcode `\^12\catcode `\_12\catcode `\%12\relax}%
\providecommand \@@startlink[1]{}%
\providecommand \@@endlink[0]{}%
\providecommand \url  [0]{\begingroup\@sanitize@url \@url }%
\providecommand \@url [1]{\endgroup\@href {#1}{\urlprefix }}%
\providecommand \urlprefix  [0]{URL }%
\providecommand \Eprint [0]{\href }%
\providecommand \doibase [0]{http://dx.doi.org/}%
\providecommand \selectlanguage [0]{\@gobble}%
\providecommand \bibinfo  [0]{\@secondoftwo}%
\providecommand \bibfield  [0]{\@secondoftwo}%
\providecommand \translation [1]{[#1]}%
\providecommand \BibitemOpen [0]{}%
\providecommand \bibitemStop [0]{}%
\providecommand \bibitemNoStop [0]{.\EOS\space}%
\providecommand \EOS [0]{\spacefactor3000\relax}%
\providecommand \BibitemShut  [1]{\csname bibitem#1\endcsname}%
\let\auto@bib@innerbib\@empty
\bibitem [{\citenamefont {Veselago}(2003)}]{Veselago2003}%
  \BibitemOpen
  \bibfield  {author} {\bibinfo {author} {\bibfnamefont {V.~G.}\ \bibnamefont
  {Veselago}},\ }\href@noop {} {\bibfield  {journal} {\bibinfo  {journal}
  {Phys.-Usp.}\ }\textbf {\bibinfo {volume} {46}},\ \bibinfo {pages} {764}
  (\bibinfo {year} {2003})}\BibitemShut {NoStop}%
\bibitem [{\citenamefont {Smith}\ \emph {et~al.}(2000)\citenamefont {Smith},
  \citenamefont {Padilla}, \citenamefont {Vier}, \citenamefont {Nemat-Nasser},\
  and\ \citenamefont {Schultz}}]{Smith2000}%
  \BibitemOpen
  \bibfield  {author} {\bibinfo {author} {\bibfnamefont {D.~R.}\ \bibnamefont
  {Smith}}, \bibinfo {author} {\bibfnamefont {W.~J.}\ \bibnamefont {Padilla}},
  \bibinfo {author} {\bibfnamefont {D.~C.}\ \bibnamefont {Vier}}, \bibinfo
  {author} {\bibfnamefont {S.~C.}\ \bibnamefont {Nemat-Nasser}}, \ and\
  \bibinfo {author} {\bibfnamefont {S.}~\bibnamefont {Schultz}},\ }\href@noop
  {} {\bibfield  {journal} {\bibinfo  {journal} {Phys. Rev. Lett.}\ }\textbf
  {\bibinfo {volume} {84}},\ \bibinfo {pages} {4184} (\bibinfo {year}
  {2000})}\BibitemShut {NoStop}%
\bibitem [{\citenamefont {Shelby}\ \emph {et~al.}(2001)\citenamefont {Shelby},
  \citenamefont {Smith},\ and\ \citenamefont {Schultz}}]{Shelby2001}%
  \BibitemOpen
  \bibfield  {author} {\bibinfo {author} {\bibfnamefont {R.~A.}\ \bibnamefont
  {Shelby}}, \bibinfo {author} {\bibfnamefont {D.~R.}\ \bibnamefont {Smith}}, \
  and\ \bibinfo {author} {\bibfnamefont {S.}~\bibnamefont {Schultz}},\
  }\href@noop {} {\bibfield  {journal} {\bibinfo  {journal} {Science}\ }\textbf
  {\bibinfo {volume} {292}},\ \bibinfo {pages} {77} (\bibinfo {year}
  {2001})}\BibitemShut {NoStop}%
\bibitem [{\citenamefont {Alu}\ and\ \citenamefont {Engheta}(2003)}]{Alu2003}%
  \BibitemOpen
  \bibfield  {author} {\bibinfo {author} {\bibfnamefont {A.}~\bibnamefont
  {Alu}}\ and\ \bibinfo {author} {\bibfnamefont {N.}~\bibnamefont {Engheta}},\
  }\href@noop {} {\bibfield  {journal} {\bibinfo  {journal} {IEEE Trans.
  Antennas Propag.}\ }\textbf {\bibinfo {volume} {51}},\ \bibinfo {pages}
  {2558} (\bibinfo {year} {2003})}\BibitemShut {NoStop}%
\bibitem [{\citenamefont {Schurig}\ \emph {et~al.}(2006)\citenamefont
  {Schurig}, \citenamefont {Mock}, \citenamefont {Justice}, \citenamefont
  {Cummer}, \citenamefont {Pendry}, \citenamefont {Starr},\ and\ \citenamefont
  {Smith}}]{Schurig2006}%
  \BibitemOpen
  \bibfield  {author} {\bibinfo {author} {\bibfnamefont {D.}~\bibnamefont
  {Schurig}}, \bibinfo {author} {\bibfnamefont {J.~J.}\ \bibnamefont {Mock}},
  \bibinfo {author} {\bibfnamefont {B.~J.}\ \bibnamefont {Justice}}, \bibinfo
  {author} {\bibfnamefont {S.~A.}\ \bibnamefont {Cummer}}, \bibinfo {author}
  {\bibfnamefont {J.~B.}\ \bibnamefont {Pendry}}, \bibinfo {author}
  {\bibfnamefont {A.~F.}\ \bibnamefont {Starr}}, \ and\ \bibinfo {author}
  {\bibfnamefont {D.~R.}\ \bibnamefont {Smith}},\ }\href@noop {} {\bibfield
  {journal} {\bibinfo  {journal} {Science}\ }\textbf {\bibinfo {volume}
  {314}},\ \bibinfo {pages} {977} (\bibinfo {year} {2006})}\BibitemShut
  {NoStop}%
\bibitem [{\citenamefont {Pendry}(2000)}]{Pendry2000}%
  \BibitemOpen
  \bibfield  {author} {\bibinfo {author} {\bibfnamefont {J.~B.}\ \bibnamefont
  {Pendry}},\ }\href@noop {} {\bibfield  {journal} {\bibinfo  {journal} {Phys.
  Rev. Lett.}\ }\textbf {\bibinfo {volume} {85}},\ \bibinfo {pages} {3966}
  (\bibinfo {year} {2000})}\BibitemShut {NoStop}%
\bibitem [{\citenamefont {Jacob}\ \emph {et~al.}(2006)\citenamefont {Jacob},
  \citenamefont {Alekseyev},\ and\ \citenamefont {Narimanov}}]{Jacob2006}%
  \BibitemOpen
  \bibfield  {author} {\bibinfo {author} {\bibfnamefont {Z.}~\bibnamefont
  {Jacob}}, \bibinfo {author} {\bibfnamefont {L.~V.}\ \bibnamefont
  {Alekseyev}}, \ and\ \bibinfo {author} {\bibfnamefont {E.}~\bibnamefont
  {Narimanov}},\ }\href@noop {} {\bibfield  {journal} {\bibinfo  {journal}
  {Opt. Express}\ }\textbf {\bibinfo {volume} {14}},\ \bibinfo {pages} {8247}
  (\bibinfo {year} {2006})}\BibitemShut {NoStop}%
\bibitem [{\citenamefont {Ricci}\ \emph {et~al.}(2005)\citenamefont {Ricci},
  \citenamefont {Orloff},\ and\ \citenamefont {Anlage}}]{Ricci2005}%
  \BibitemOpen
  \bibfield  {author} {\bibinfo {author} {\bibfnamefont {M.}~\bibnamefont
  {Ricci}}, \bibinfo {author} {\bibfnamefont {N.}~\bibnamefont {Orloff}}, \
  and\ \bibinfo {author} {\bibfnamefont {S.~M.}\ \bibnamefont {Anlage}},\
  }\href@noop {} {\bibfield  {journal} {\bibinfo  {journal} {Appl. Phys.
  Lett.}\ }\textbf {\bibinfo {volume} {87}},\ \bibinfo {pages} {034102}
  (\bibinfo {year} {2005})}\BibitemShut {NoStop}%
\bibitem [{\citenamefont {Ricci}\ and\ \citenamefont
  {Anlage}(2006)}]{Ricci2006}%
  \BibitemOpen
  \bibfield  {author} {\bibinfo {author} {\bibfnamefont {M.~C.}\ \bibnamefont
  {Ricci}}\ and\ \bibinfo {author} {\bibfnamefont {S.~M.}\ \bibnamefont
  {Anlage}},\ }\href@noop {} {\bibfield  {journal} {\bibinfo  {journal} {Appl.
  Phys. Lett.}\ }\textbf {\bibinfo {volume} {88}},\ \bibinfo {pages} {264102}
  (\bibinfo {year} {2006})}\BibitemShut {NoStop}%
\bibitem [{\citenamefont {Ricci}\ \emph {et~al.}(2007)\citenamefont {Ricci},
  \citenamefont {Xu}, \citenamefont {Prozorov}, \citenamefont {Zhuravel},
  \citenamefont {Ustinov},\ and\ \citenamefont {Anlage}}]{Ricci2007a}%
  \BibitemOpen
  \bibfield  {author} {\bibinfo {author} {\bibfnamefont {M.~C.}\ \bibnamefont
  {Ricci}}, \bibinfo {author} {\bibfnamefont {H.}~\bibnamefont {Xu}}, \bibinfo
  {author} {\bibfnamefont {R.}~\bibnamefont {Prozorov}}, \bibinfo {author}
  {\bibfnamefont {A.~P.}\ \bibnamefont {Zhuravel}}, \bibinfo {author}
  {\bibfnamefont {A.~V.}\ \bibnamefont {Ustinov}}, \ and\ \bibinfo {author}
  {\bibfnamefont {S.~M.}\ \bibnamefont {Anlage}},\ }\href@noop {} {\bibfield
  {journal} {\bibinfo  {journal} {IEEE Trans. Appl. Supercond.}\ }\textbf
  {\bibinfo {volume} {17}},\ \bibinfo {pages} {918} (\bibinfo {year}
  {2007})}\BibitemShut {NoStop}%
\bibitem [{\citenamefont {Fedotov}\ \emph
  {et~al.}(2010{\natexlab{a}})\citenamefont {Fedotov}, \citenamefont
  {Tsiatmas}, \citenamefont {Shi}, \citenamefont {Buckingham}, \citenamefont
  {de~Groot}, \citenamefont {Chen}, \citenamefont {Wang},\ and\ \citenamefont
  {Zheludev}}]{Fedotov2010OE}%
  \BibitemOpen
  \bibfield  {author} {\bibinfo {author} {\bibfnamefont {V.}~\bibnamefont
  {Fedotov}}, \bibinfo {author} {\bibfnamefont {A.}~\bibnamefont {Tsiatmas}},
  \bibinfo {author} {\bibfnamefont {J.~H.}\ \bibnamefont {Shi}}, \bibinfo
  {author} {\bibfnamefont {R.}~\bibnamefont {Buckingham}}, \bibinfo {author}
  {\bibfnamefont {P.}~\bibnamefont {de~Groot}}, \bibinfo {author}
  {\bibfnamefont {Y.}~\bibnamefont {Chen}}, \bibinfo {author} {\bibfnamefont
  {S.}~\bibnamefont {Wang}}, \ and\ \bibinfo {author} {\bibfnamefont
  {N.}~\bibnamefont {Zheludev}},\ }\href {\doibase 10.1364/OE.18.009015}
  {\bibfield  {journal} {\bibinfo  {journal} {Opt. Express}\ }\textbf {\bibinfo
  {volume} {18}},\ \bibinfo {pages} {9015} (\bibinfo {year}
  {2010}{\natexlab{a}})}\BibitemShut {NoStop}%
\bibitem [{\citenamefont {Savinov}\ \emph {et~al.}(2012)\citenamefont
  {Savinov}, \citenamefont {Tsiatmas}, \citenamefont {Buckingham},
  \citenamefont {Fedotov}, \citenamefont {de~Groot},\ and\ \citenamefont
  {Zheludev}}]{Savinov2012}%
  \BibitemOpen
  \bibfield  {author} {\bibinfo {author} {\bibfnamefont {V.}~\bibnamefont
  {Savinov}}, \bibinfo {author} {\bibfnamefont {A.}~\bibnamefont {Tsiatmas}},
  \bibinfo {author} {\bibfnamefont {A.~R.}\ \bibnamefont {Buckingham}},
  \bibinfo {author} {\bibfnamefont {V.~A.}\ \bibnamefont {Fedotov}}, \bibinfo
  {author} {\bibfnamefont {P.~A.~J.}\ \bibnamefont {de~Groot}}, \ and\ \bibinfo
  {author} {\bibfnamefont {N.~I.}\ \bibnamefont {Zheludev}},\ }\href {\doibase
  10.1038/srep00450} {\bibfield  {journal} {\bibinfo  {journal} {Sci Rep}\
  }\textbf {\bibinfo {volume} {2}},\ \bibinfo {pages} {450} (\bibinfo {year}
  {2012})}\BibitemShut {NoStop}%
\bibitem [{\citenamefont {Trepanier}\ \emph {et~al.}(2013)\citenamefont
  {Trepanier}, \citenamefont {Zhang}, \citenamefont {Mukhanov},\ and\
  \citenamefont {Anlage}}]{Trepanier2013}%
  \BibitemOpen
  \bibfield  {author} {\bibinfo {author} {\bibfnamefont {M.}~\bibnamefont
  {Trepanier}}, \bibinfo {author} {\bibfnamefont {D.}~\bibnamefont {Zhang}},
  \bibinfo {author} {\bibfnamefont {O.}~\bibnamefont {Mukhanov}}, \ and\
  \bibinfo {author} {\bibfnamefont {S.~M.}\ \bibnamefont {Anlage}},\
  }\href@noop {} {\bibfield  {journal} {\bibinfo  {journal} {Phys. Rev. X}\
  }\textbf {\bibinfo {volume} {3}},\ \bibinfo {pages} {041029} (\bibinfo {year}
  {2013})}\BibitemShut {NoStop}%
\bibitem [{\citenamefont {Du}\ \emph {et~al.}(2006)\citenamefont {Du},
  \citenamefont {Chen},\ and\ \citenamefont {Li}}]{Du2006}%
  \BibitemOpen
  \bibfield  {author} {\bibinfo {author} {\bibfnamefont {C.}~\bibnamefont
  {Du}}, \bibinfo {author} {\bibfnamefont {H.}~\bibnamefont {Chen}}, \ and\
  \bibinfo {author} {\bibfnamefont {S.}~\bibnamefont {Li}},\ }\href@noop {}
  {\bibfield  {journal} {\bibinfo  {journal} {Phys. Rev. B}\ }\textbf {\bibinfo
  {volume} {74}},\ \bibinfo {pages} {113105} (\bibinfo {year}
  {2006})}\BibitemShut {NoStop}%
\bibitem [{\citenamefont {Lazarides}\ and\ \citenamefont
  {Tsironis}(2007)}]{Lazarides2007}%
  \BibitemOpen
  \bibfield  {author} {\bibinfo {author} {\bibfnamefont {N.}~\bibnamefont
  {Lazarides}}\ and\ \bibinfo {author} {\bibfnamefont {G.~P.}\ \bibnamefont
  {Tsironis}},\ }\href@noop {} {\bibfield  {journal} {\bibinfo  {journal}
  {Appl. Phys. Lett.}\ }\textbf {\bibinfo {volume} {90}},\ \bibinfo {pages}
  {163501} (\bibinfo {year} {2007})}\BibitemShut {NoStop}%
\bibitem [{\citenamefont {Caputo}\ \emph {et~al.}(2012)\citenamefont {Caputo},
  \citenamefont {Gabitov},\ and\ \citenamefont {Maimistov}}]{Caputo2012}%
  \BibitemOpen
  \bibfield  {author} {\bibinfo {author} {\bibfnamefont {J.-G.}\ \bibnamefont
  {Caputo}}, \bibinfo {author} {\bibfnamefont {I.}~\bibnamefont {Gabitov}}, \
  and\ \bibinfo {author} {\bibfnamefont {A.}~\bibnamefont {Maimistov}},\
  }\href@noop {} {\bibfield  {journal} {\bibinfo  {journal} {Phys. Rev. B}\
  }\textbf {\bibinfo {volume} {85}},\ \bibinfo {pages} {205446} (\bibinfo
  {year} {2012})}\BibitemShut {NoStop}%
\bibitem [{\citenamefont {Jung}\ \emph {et~al.}(2013)\citenamefont {Jung},
  \citenamefont {Butz}, \citenamefont {Shitov},\ and\ \citenamefont
  {Ustinov}}]{Jung2013}%
  \BibitemOpen
  \bibfield  {author} {\bibinfo {author} {\bibfnamefont {P.}~\bibnamefont
  {Jung}}, \bibinfo {author} {\bibfnamefont {S.}~\bibnamefont {Butz}}, \bibinfo
  {author} {\bibfnamefont {S.~V.}\ \bibnamefont {Shitov}}, \ and\ \bibinfo
  {author} {\bibfnamefont {A.~V.}\ \bibnamefont {Ustinov}},\ }\href@noop {}
  {\bibfield  {journal} {\bibinfo  {journal} {Appl. Phys. Lett.}\ }\textbf
  {\bibinfo {volume} {102}},\ \bibinfo {pages} {062601} (\bibinfo {year}
  {2013})}\BibitemShut {NoStop}%
\bibitem [{\citenamefont {Butz}\ \emph
  {et~al.}(2013{\natexlab{a}})\citenamefont {Butz}, \citenamefont {Jung},
  \citenamefont {Filippenko}, \citenamefont {Koshelets},\ and\ \citenamefont
  {Ustinov}}]{Butz2013}%
  \BibitemOpen
  \bibfield  {author} {\bibinfo {author} {\bibfnamefont {S.}~\bibnamefont
  {Butz}}, \bibinfo {author} {\bibfnamefont {P.}~\bibnamefont {Jung}}, \bibinfo
  {author} {\bibfnamefont {L.~V.}\ \bibnamefont {Filippenko}}, \bibinfo
  {author} {\bibfnamefont {V.~P.}\ \bibnamefont {Koshelets}}, \ and\ \bibinfo
  {author} {\bibfnamefont {A.~V.}\ \bibnamefont {Ustinov}},\ }\href@noop {}
  {\bibfield  {journal} {\bibinfo  {journal} {Supercond. Sci. Technol.}\
  }\textbf {\bibinfo {volume} {26}},\ \bibinfo {pages} {094004} (\bibinfo
  {year} {2013}{\natexlab{a}})}\BibitemShut {NoStop}%
\bibitem [{\citenamefont {Butz}\ \emph
  {et~al.}(2013{\natexlab{b}})\citenamefont {Butz}, \citenamefont {Jung},
  \citenamefont {Filippenko}, \citenamefont {Koshelets},\ and\ \citenamefont
  {Ustinov}}]{Butz2013a}%
  \BibitemOpen
  \bibfield  {author} {\bibinfo {author} {\bibfnamefont {S.}~\bibnamefont
  {Butz}}, \bibinfo {author} {\bibfnamefont {P.}~\bibnamefont {Jung}}, \bibinfo
  {author} {\bibfnamefont {L.~V.}\ \bibnamefont {Filippenko}}, \bibinfo
  {author} {\bibfnamefont {V.~P.}\ \bibnamefont {Koshelets}}, \ and\ \bibinfo
  {author} {\bibfnamefont {A.~V.}\ \bibnamefont {Ustinov}},\ }\href@noop {}
  {\bibfield  {journal} {\bibinfo  {journal} {Opt Express}\ }\textbf {\bibinfo
  {volume} {21}},\ \bibinfo {pages} {22540} (\bibinfo {year}
  {2013}{\natexlab{b}})}\BibitemShut {NoStop}%
\bibitem [{\citenamefont {Jung}\ \emph {et~al.}(2014)\citenamefont {Jung},
  \citenamefont {Butz}, \citenamefont {Marthaler}, \citenamefont {Fistul},
  \citenamefont {Lepp{\"a}kangas}, \citenamefont {Koshelets},\ and\
  \citenamefont {Ustinov}}]{Jung2014}%
  \BibitemOpen
  \bibfield  {author} {\bibinfo {author} {\bibfnamefont {P.}~\bibnamefont
  {Jung}}, \bibinfo {author} {\bibfnamefont {S.}~\bibnamefont {Butz}}, \bibinfo
  {author} {\bibfnamefont {M.}~\bibnamefont {Marthaler}}, \bibinfo {author}
  {\bibfnamefont {M.~V.}\ \bibnamefont {Fistul}}, \bibinfo {author}
  {\bibfnamefont {J.}~\bibnamefont {Lepp{\"a}kangas}}, \bibinfo {author}
  {\bibfnamefont {V.~P.}\ \bibnamefont {Koshelets}}, \ and\ \bibinfo {author}
  {\bibfnamefont {A.~V.}\ \bibnamefont {Ustinov}},\ }\href {\doibase
  10.1038/ncomms4730} {\bibfield  {journal} {\bibinfo  {journal} {Nat Commun}\
  }\textbf {\bibinfo {volume} {5}},\ \bibinfo {pages} {3730} (\bibinfo {year}
  {2014})}\BibitemShut {NoStop}%
\bibitem [{\citenamefont {Lazarides}\ and\ \citenamefont
  {Tsironis}(2013)}]{Lazarides2013}%
  \BibitemOpen
  \bibfield  {author} {\bibinfo {author} {\bibfnamefont {N.}~\bibnamefont
  {Lazarides}}\ and\ \bibinfo {author} {\bibfnamefont {G.~P.}\ \bibnamefont
  {Tsironis}},\ }\href@noop {} {\bibfield  {journal} {\bibinfo  {journal}
  {Supercond. Sci. Technol.}\ }\textbf {\bibinfo {volume} {26}},\ \bibinfo
  {pages} {084006} (\bibinfo {year} {2013})}\BibitemShut {NoStop}%
\bibitem [{\citenamefont {Lazarides}\ and\ \citenamefont
  {Tsironis}(2010)}]{Lazarides2010}%
  \BibitemOpen
  \bibfield  {author} {\bibinfo {author} {\bibfnamefont {N.}~\bibnamefont
  {Lazarides}}\ and\ \bibinfo {author} {\bibfnamefont {G.}~\bibnamefont
  {Tsironis}},\ }\href@noop {} {\bibfield  {journal} {\bibinfo  {journal}
  {Phys. Lett. A}\ }\textbf {\bibinfo {volume} {374}},\ \bibinfo {pages} {2179}
  (\bibinfo {year} {2010})}\BibitemShut {NoStop}%
\bibitem [{\citenamefont {Tsironis}\ \emph {et~al.}(2014)\citenamefont
  {Tsironis}, \citenamefont {Lazarides},\ and\ \citenamefont
  {Margaris}}]{Tsironis2014}%
  \BibitemOpen
  \bibfield  {author} {\bibinfo {author} {\bibfnamefont {G.~P.}\ \bibnamefont
  {Tsironis}}, \bibinfo {author} {\bibfnamefont {N.}~\bibnamefont {Lazarides}},
  \ and\ \bibinfo {author} {\bibfnamefont {I.}~\bibnamefont {Margaris}},\
  }\href@noop {} {\bibfield  {journal} {\bibinfo  {journal} {Appl. Phys.
  Lett.}\ }\textbf {\bibinfo {volume} {117}},\ \bibinfo {pages} {579} (\bibinfo
  {year} {2014})}\BibitemShut {NoStop}%
\bibitem [{\citenamefont {Lazarides}\ and\ \citenamefont
  {Tsironis}(2015)}]{Lazarides2015a}%
  \BibitemOpen
  \bibfield  {author} {\bibinfo {author} {\bibfnamefont {N.}~\bibnamefont
  {Lazarides}}\ and\ \bibinfo {author} {\bibfnamefont {G.~P.}\ \bibnamefont
  {Tsironis}},\ }in\ \href@noop {} {\emph {\bibinfo {booktitle} {Nonlinear,
  Tunable and Active Metamaterials}}}\ (\bibinfo  {publisher} {Springer
  International Publishing},\ \bibinfo {year} {2015})\ pp.\ \bibinfo {pages}
  {281--301}\BibitemShut {NoStop}%
\bibitem [{\citenamefont {Lazarides}\ \emph {et~al.}(2015)\citenamefont
  {Lazarides}, \citenamefont {Neofotistos},\ and\ \citenamefont
  {Tsironis}}]{Lazarides2015}%
  \BibitemOpen
  \bibfield  {author} {\bibinfo {author} {\bibfnamefont {N.}~\bibnamefont
  {Lazarides}}, \bibinfo {author} {\bibfnamefont {G.}~\bibnamefont
  {Neofotistos}}, \ and\ \bibinfo {author} {\bibfnamefont {G.~P.}\ \bibnamefont
  {Tsironis}},\ }\href@noop {} {\bibfield  {journal} {\bibinfo  {journal}
  {Phys. Rev. B}\ }\textbf {\bibinfo {volume} {91}},\ \bibinfo {pages} {054303}
  (\bibinfo {year} {2015})}\BibitemShut {NoStop}%
\bibitem [{\citenamefont {Zhang}\ \emph {et~al.}(2016)\citenamefont {Zhang},
  \citenamefont {Trepanier}, \citenamefont {Antonsen}, \citenamefont {Ott},\
  and\ \citenamefont {Anlage}}]{Zhang2016}%
  \BibitemOpen
  \bibfield  {author} {\bibinfo {author} {\bibfnamefont {D.}~\bibnamefont
  {Zhang}}, \bibinfo {author} {\bibfnamefont {M.}~\bibnamefont {Trepanier}},
  \bibinfo {author} {\bibfnamefont {T.}~\bibnamefont {Antonsen}}, \bibinfo
  {author} {\bibfnamefont {E.}~\bibnamefont {Ott}}, \ and\ \bibinfo {author}
  {\bibfnamefont {S.~M.}\ \bibnamefont {Anlage}},\ }\href {\doibase
  10.1103/PhysRevB.94.174507} {\bibfield  {journal} {\bibinfo  {journal} {Phys.
  Rev. B}\ }\textbf {\bibinfo {volume} {94}},\ \bibinfo {pages} {174507}
  (\bibinfo {year} {2016})}\BibitemShut {NoStop}%
\bibitem [{\citenamefont {Tsang}\ \emph {et~al.}(1991)\citenamefont {Tsang},
  \citenamefont {Mirollo}, \citenamefont {Strogatz},\ and\ \citenamefont
  {Wiesenfeld}}]{Tsang1991}%
  \BibitemOpen
  \bibfield  {author} {\bibinfo {author} {\bibfnamefont {K.~Y.}\ \bibnamefont
  {Tsang}}, \bibinfo {author} {\bibfnamefont {R.~E.}\ \bibnamefont {Mirollo}},
  \bibinfo {author} {\bibfnamefont {S.~H.}\ \bibnamefont {Strogatz}}, \ and\
  \bibinfo {author} {\bibfnamefont {K.}~\bibnamefont {Wiesenfeld}},\
  }\href@noop {} {\bibfield  {journal} {\bibinfo  {journal} {Physica D}\
  }\textbf {\bibinfo {volume} {48}},\ \bibinfo {pages} {102} (\bibinfo {year}
  {1991})}\BibitemShut {NoStop}%
\bibitem [{\citenamefont {Acebr{\'o}n}\ \emph {et~al.}(2005)\citenamefont
  {Acebr{\'o}n}, \citenamefont {Bonilla}, \citenamefont {Vicente},
  \citenamefont {Ritort},\ and\ \citenamefont {Spigler}}]{Acebron2005}%
  \BibitemOpen
  \bibfield  {author} {\bibinfo {author} {\bibfnamefont {J.~A.}\ \bibnamefont
  {Acebr{\'o}n}}, \bibinfo {author} {\bibfnamefont {L.~L.}\ \bibnamefont
  {Bonilla}}, \bibinfo {author} {\bibfnamefont {C.~J.~P.}\ \bibnamefont
  {Vicente}}, \bibinfo {author} {\bibfnamefont {F.}~\bibnamefont {Ritort}}, \
  and\ \bibinfo {author} {\bibfnamefont {R.}~\bibnamefont {Spigler}},\
  }\href@noop {} {\bibfield  {journal} {\bibinfo  {journal} {Rev. Mod. Phys.}\
  }\textbf {\bibinfo {volume} {77}},\ \bibinfo {pages} {137} (\bibinfo {year}
  {2005})}\BibitemShut {NoStop}%
\bibitem [{\citenamefont {Marvel}\ and\ \citenamefont
  {Strogatz}(2009)}]{Marvel2009}%
  \BibitemOpen
  \bibfield  {author} {\bibinfo {author} {\bibfnamefont {S.~A.}\ \bibnamefont
  {Marvel}}\ and\ \bibinfo {author} {\bibfnamefont {S.~H.}\ \bibnamefont
  {Strogatz}},\ }\href {\doibase 10.1063/1.3087132} {\bibfield  {journal}
  {\bibinfo  {journal} {Chaos}\ }\textbf {\bibinfo {volume} {19}},\ \bibinfo
  {pages} {013132} (\bibinfo {year} {2009})}\BibitemShut {NoStop}%
\bibitem [{\citenamefont {Fedotov}\ \emph
  {et~al.}(2010{\natexlab{b}})\citenamefont {Fedotov}, \citenamefont
  {Papasimakis}, \citenamefont {Plum}, \citenamefont {Bitzer}, \citenamefont
  {Walther}, \citenamefont {Kuo}, \citenamefont {Tsai},\ and\ \citenamefont
  {Zheludev}}]{Fedotov2010}%
  \BibitemOpen
  \bibfield  {author} {\bibinfo {author} {\bibfnamefont {V.~A.}\ \bibnamefont
  {Fedotov}}, \bibinfo {author} {\bibfnamefont {N.}~\bibnamefont
  {Papasimakis}}, \bibinfo {author} {\bibfnamefont {E.}~\bibnamefont {Plum}},
  \bibinfo {author} {\bibfnamefont {A.}~\bibnamefont {Bitzer}}, \bibinfo
  {author} {\bibfnamefont {M.}~\bibnamefont {Walther}}, \bibinfo {author}
  {\bibfnamefont {P.}~\bibnamefont {Kuo}}, \bibinfo {author} {\bibfnamefont
  {D.~P.}\ \bibnamefont {Tsai}}, \ and\ \bibinfo {author} {\bibfnamefont
  {N.~I.}\ \bibnamefont {Zheludev}},\ }\href@noop {} {\bibfield  {journal}
  {\bibinfo  {journal} {Phys. Rev. Lett.}\ }\textbf {\bibinfo {volume} {104}},\
  \bibinfo {pages} {223901} (\bibinfo {year} {2010}{\natexlab{b}})}\BibitemShut
  {NoStop}%
\bibitem [{\citenamefont {Jenkins}\ and\ \citenamefont
  {Ruostekoski}(2012)}]{Jenkins2012}%
  \BibitemOpen
  \bibfield  {author} {\bibinfo {author} {\bibfnamefont {S.~D.}\ \bibnamefont
  {Jenkins}}\ and\ \bibinfo {author} {\bibfnamefont {J.}~\bibnamefont
  {Ruostekoski}},\ }\href@noop {} {\bibfield  {journal} {\bibinfo  {journal}
  {Phys. Rev. B}\ }\textbf {\bibinfo {volume} {86}},\ \bibinfo {pages} {205128}
  (\bibinfo {year} {2012})}\BibitemShut {NoStop}%
\bibitem [{\citenamefont {Jenkins}\ and\ \citenamefont
  {Ruostekoski}(2013)}]{Jenkins2013}%
  \BibitemOpen
  \bibfield  {author} {\bibinfo {author} {\bibfnamefont {S.~D.}\ \bibnamefont
  {Jenkins}}\ and\ \bibinfo {author} {\bibfnamefont {J.}~\bibnamefont
  {Ruostekoski}},\ }\href {\doibase 10.1103/PhysRevLett.111.147401} {\bibfield
  {journal} {\bibinfo  {journal} {Phys. Rev. Lett.}\ }\textbf {\bibinfo
  {volume} {111}},\ \bibinfo {pages} {147401} (\bibinfo {year}
  {2013})}\BibitemShut {NoStop}%
\bibitem [{\citenamefont {Zhang}\ \emph {et~al.}(2015)\citenamefont {Zhang},
  \citenamefont {Trepanier}, \citenamefont {Mukhanov},\ and\ \citenamefont
  {Anlage}}]{Zhang2015}%
  \BibitemOpen
  \bibfield  {author} {\bibinfo {author} {\bibfnamefont {D.}~\bibnamefont
  {Zhang}}, \bibinfo {author} {\bibfnamefont {M.}~\bibnamefont {Trepanier}},
  \bibinfo {author} {\bibfnamefont {O.}~\bibnamefont {Mukhanov}}, \ and\
  \bibinfo {author} {\bibfnamefont {S.~M.}\ \bibnamefont {Anlage}},\
  }\href@noop {} {\bibfield  {journal} {\bibinfo  {journal} {Phys. Rev. X}\
  }\textbf {\bibinfo {volume} {5}},\ \bibinfo {pages} {041045} (\bibinfo {year}
  {2015})}\BibitemShut {NoStop}%
\bibitem [{\citenamefont {Koshelets}\ \emph {et~al.}(1989)\citenamefont
  {Koshelets}, \citenamefont {Matlashov}, \citenamefont {Serpuchenko},
  \citenamefont {Filippenko},\ and\ \citenamefont {Zhuravlev}}]{Koshelets1989}%
  \BibitemOpen
  \bibfield  {author} {\bibinfo {author} {\bibfnamefont {V.~P.}\ \bibnamefont
  {Koshelets}}, \bibinfo {author} {\bibfnamefont {A.~N.}\ \bibnamefont
  {Matlashov}}, \bibinfo {author} {\bibfnamefont {I.~L.}\ \bibnamefont
  {Serpuchenko}}, \bibinfo {author} {\bibfnamefont {L.~V.}\ \bibnamefont
  {Filippenko}}, \ and\ \bibinfo {author} {\bibfnamefont {Y.~E.}\ \bibnamefont
  {Zhuravlev}},\ }\href@noop {} {\bibfield  {journal} {\bibinfo  {journal}
  {IEEE Trans. Magn.}\ }\textbf {\bibinfo {volume} {25}},\ \bibinfo {pages}
  {1182} (\bibinfo {year} {1989})}\BibitemShut {NoStop}%
\bibitem [{\citenamefont {Koshelets}\ \emph {et~al.}(1991)\citenamefont
  {Koshelets}, \citenamefont {Kovtonyuk}, \citenamefont {Serpuchenko},
  \citenamefont {Filippenko},\ and\ \citenamefont {Shchukin}}]{Koshelets1991}%
  \BibitemOpen
  \bibfield  {author} {\bibinfo {author} {\bibfnamefont {V.~P.}\ \bibnamefont
  {Koshelets}}, \bibinfo {author} {\bibfnamefont {S.~A.}\ \bibnamefont
  {Kovtonyuk}}, \bibinfo {author} {\bibfnamefont {I.~L.}\ \bibnamefont
  {Serpuchenko}}, \bibinfo {author} {\bibfnamefont {L.~V.}\ \bibnamefont
  {Filippenko}}, \ and\ \bibinfo {author} {\bibfnamefont {A.~V.}\ \bibnamefont
  {Shchukin}},\ }\href@noop {} {\bibfield  {journal} {\bibinfo  {journal} {IEEE
  Trans. Magn.}\ }\textbf {\bibinfo {volume} {27}},\ \bibinfo {pages} {3141}
  (\bibinfo {year} {1991})}\BibitemShut {NoStop}%
\bibitem [{\citenamefont {Filippenko}\ \emph {et~al.}(2001)\citenamefont
  {Filippenko}, \citenamefont {Shitov}, \citenamefont {Dmitriev}, \citenamefont
  {Ermakov}, \citenamefont {Koshelets},\ and\ \citenamefont
  {Gao}}]{Filippenko2001}%
  \BibitemOpen
  \bibfield  {author} {\bibinfo {author} {\bibfnamefont {L.~V.}\ \bibnamefont
  {Filippenko}}, \bibinfo {author} {\bibfnamefont {S.~V.}\ \bibnamefont
  {Shitov}}, \bibinfo {author} {\bibfnamefont {P.~N.}\ \bibnamefont
  {Dmitriev}}, \bibinfo {author} {\bibfnamefont {A.~B.}\ \bibnamefont
  {Ermakov}}, \bibinfo {author} {\bibfnamefont {V.~P.}\ \bibnamefont
  {Koshelets}}, \ and\ \bibinfo {author} {\bibfnamefont {J.-R.}\ \bibnamefont
  {Gao}},\ }\href@noop {} {\bibfield  {journal} {\bibinfo  {journal} {IEEE
  Trans. Appl. Supercond.}\ }\textbf {\bibinfo {volume} {11}},\ \bibinfo
  {pages} {816} (\bibinfo {year} {2001})}\BibitemShut {NoStop}%
\bibitem [{\citenamefont {Research}()}]{Whiteley}%
  \BibitemOpen
  \bibfield  {author} {\bibinfo {author} {\bibfnamefont {W.}~\bibnamefont
  {Research}},\ }\href {http://www.wrcad.com/} {}\bibinfo {howpublished}
  {\url{http://www.wrcad.com/}}\BibitemShut {NoStop}%
\bibitem [{\citenamefont {Yohannes}\ \emph {et~al.}(2005)\citenamefont
  {Yohannes}, \citenamefont {Sarwana}, \citenamefont {Tolpygo}, \citenamefont
  {Sahu}, \citenamefont {Polyakov},\ and\ \citenamefont
  {Semenov}}]{Yohannes2005}%
  \BibitemOpen
  \bibfield  {author} {\bibinfo {author} {\bibfnamefont {D.}~\bibnamefont
  {Yohannes}}, \bibinfo {author} {\bibfnamefont {S.}~\bibnamefont {Sarwana}},
  \bibinfo {author} {\bibfnamefont {S.~K.}\ \bibnamefont {Tolpygo}}, \bibinfo
  {author} {\bibfnamefont {A.}~\bibnamefont {Sahu}}, \bibinfo {author}
  {\bibfnamefont {Y.~A.}\ \bibnamefont {Polyakov}}, \ and\ \bibinfo {author}
  {\bibfnamefont {V.~K.}\ \bibnamefont {Semenov}},\ }\href@noop {} {\bibfield
  {journal} {\bibinfo  {journal} {IEEE Trans. Appl. Supercond.}\ }\textbf
  {\bibinfo {volume} {15}},\ \bibinfo {pages} {90} (\bibinfo {year}
  {2005})}\BibitemShut {NoStop}%
\bibitem [{\citenamefont {Yohannes}\ \emph {et~al.}(2007)\citenamefont
  {Yohannes}, \citenamefont {Kirichenko}, \citenamefont {Sarwana},\ and\
  \citenamefont {Tolpygo}}]{Yohannes2007}%
  \BibitemOpen
  \bibfield  {author} {\bibinfo {author} {\bibfnamefont {D.}~\bibnamefont
  {Yohannes}}, \bibinfo {author} {\bibfnamefont {A.}~\bibnamefont
  {Kirichenko}}, \bibinfo {author} {\bibfnamefont {S.}~\bibnamefont {Sarwana}},
  \ and\ \bibinfo {author} {\bibfnamefont {S.~K.}\ \bibnamefont {Tolpygo}},\
  }\href@noop {} {\bibfield  {journal} {\bibinfo  {journal} {IEEE Trans. Appl.
  Supercond.}\ }\textbf {\bibinfo {volume} {17}},\ \bibinfo {pages} {181}
  (\bibinfo {year} {2007})}\BibitemShut {NoStop}%
\bibitem [{\citenamefont {Tolpygo}\ and\ \citenamefont
  {Amparo}(2010)}]{Tolpygo2010}%
  \BibitemOpen
  \bibfield  {author} {\bibinfo {author} {\bibfnamefont {S.~K.}\ \bibnamefont
  {Tolpygo}}\ and\ \bibinfo {author} {\bibfnamefont {D.}~\bibnamefont
  {Amparo}},\ }\href@noop {} {\bibfield  {journal} {\bibinfo  {journal}
  {Supercond. Sci. Technol.}\ }\textbf {\bibinfo {volume} {23}},\ \bibinfo
  {pages} {034024} (\bibinfo {year} {2010})}\BibitemShut {NoStop}%
\end{thebibliography}%
\end{document}